\documentclass[twocolumn,aps,prl,superscriptaddress,amsmath,amssymb,preprintnumbers,floatfix]{revtex4-2}

\usepackage[utf8]{inputenc}
\usepackage{amsmath,amssymb}
\usepackage{mathtools}
\usepackage{bm}
\usepackage{times}
\usepackage{graphicx,xcolor,soul,sidecap}
\usepackage{xr}
\usepackage[colorlinks,bookmarks=false,citecolor=blue,linkcolor=black,urlcolor=blue,hyperfootnotes=true]{hyperref}
\usepackage[capitalize]{cleveref}
\makeatother
\usepackage[separate-uncertainty = true, multi-part-units=single]{siunitx} 
\usepackage[version=4]{mhchem}

\begin{document}


\title{Imaging the suppression of ferromagnetism in LaMnO$_3$ by metallic overlayers} 

\author{Bart {Folkers}}
\thanks{These authors contributed equally to this work}
\affiliation{MESA+ Institute for Nanotechnology, University of Twente, 7500 AE Enschede, The Netherlands}
\author{Thies {Jansen}}
\thanks{These authors contributed equally to this work}
\affiliation{MESA+ Institute for Nanotechnology, University of Twente, 7500 AE Enschede, The Netherlands}
\author{Thijs J. {Roskamp}}
\thanks{These authors contributed equally to this work}
\affiliation{MESA+ Institute for Nanotechnology, University of Twente, 7500 AE Enschede, The Netherlands}
\author{Pim {Reith}}
\affiliation{MESA+ Institute for Nanotechnology, University of Twente, 7500 AE Enschede, The Netherlands}
\author{André {Timmermans}}
\affiliation{MESA+ Institute for Nanotechnology, University of Twente, 7500 AE Enschede, The Netherlands}
\author{Daen {Jannis}}
\affiliation{Electron Microscopy for Materials Research (EMAT), Department of Physics, University of Antwerp, BE-2020 Antwerpen, Belgium}
\author{Nicolas {Gauquelin}}
\affiliation{Electron Microscopy for Materials Research (EMAT), Department of Physics, University of Antwerp, BE-2020 Antwerpen, Belgium}
\affiliation{NANOlab Center of Excellence, University of Antwerp, BE-2020 Antwerpen, Belgium}
\author{Johan {Verbeeck}}
\affiliation{Electron Microscopy for Materials Research (EMAT), Department of Physics, University of Antwerp, BE-2020 Antwerpen, Belgium}
\author{Hans {Hilgenkamp}}
\email{j.w.m.hilgenkamp@utwente.nl}
\affiliation{MESA+ Institute for Nanotechnology, University of Twente, 7500 AE Enschede, The Netherlands}
\author{Carlos M. M. {Rosário}}
\affiliation{MESA+ Institute for Nanotechnology, University of Twente, 7500 AE Enschede, The Netherlands}
\affiliation{INL - International Iberian Nanotechnology Laboratory, 4715-330 Braga, Portugal}

\date{\today}

\begin{abstract}
LaMnO$_3$ (LMO) thin films epitaxially grown on SrTiO$_3$ (STO) usually exhibit ferromagnetism above a critical layer thickness. We report the use of scanning SQUID microscopy (SSM) to study the suppression of the ferromagnetism in STO/LMO/metal structures. By partially covering the LMO surface with a metallic layer, both covered and uncovered LMO regions can be studied simultaneously. While Au does not significantly influence the ferromagnetic order of the underlying LMO film, a thin Ti layer induces a strong suppression of the ferromagnetism, over tens of nanometers, which increases with time on a timescale of days. Detailed EELS analysis of the Ti-LaMnO$_3$ interface reveals \textcolor{black}{the presence of Mn$^{2+}$ and} an evolution of the Ti valence state from Ti$^0$ to Ti$^{4+}$ over approximately 5 nanometers. Furthermore, we demonstrate that by patterning Ti/Au overlayers, we can locally suppress the ferromagnetism and define ferromagnetic structures down to sub-micrometer scales.
\end{abstract}


\maketitle
Complex oxide thin films exhibit remarkable versatility, making them suitable for many potential applications in electronics, spintronics and catalysis, among other fields \cite{Bibes2011, Trier2022, Freund2008, Coll2019}. Their tunability allows for precise control and manipulation of their properties, for example by the background pressure during growth \cite{Marton2010, Kim2010, Koster2020}, the incorporation of the oxide in a heterostructure \cite{Doennig2015, Zubko2011, Huang2018, Liao2016}, the modulation of strain \cite{Hwang2019, Xu2020_STO}, doping, or selecting the thickness \cite{Shen2015, Wang2015}. LaMnO$_3$ (LMO) is a good example of such a complex oxide with versatile properties, in particular its magnetic properties. LMO is an antiferromagnetic insulator in the bulk, but becomes ferromagnetic when grown epitaxially on SrTiO$_3$ (STO) under the right oxygen background conditions \cite{Kim2010}. However, the ferromagnetism only arises above a critical thickness of 6 unit cells \cite{Wang2015, Chen2017}.

For a thin film of a complex oxide, such as LMO, to be incorporated into an electronic device, it is imperative to contact it with conventional metals. When the electronic properties are studied, it is common to contact the complex oxides with a Ti/Au layer, where Ti serves as an adhesion layer \cite{Vogt1994,Todeschini2017}. However, it is widely known that Ti has a high oxygen affinity, easily oxidizing in contact with oxygen from the environment \cite{Liu1988, Vogt1994} or with oxygen-rich materials, in the right thermodynamic conditions \cite{Kim2004,Guo2014}. Homonnay et al. \cite{Homonnay2015} also showed that a Ti overlayer on \ce{La_{0.7}Sr_{0.3}MnO3} (LSMO) can have a significant impact on the oxide stoichiometry and crystal structure. This resulted in a global disappearance of ferromagnetism as measured by SQUID vibrating sample magnetometry. Although it is clear that diffusion of oxygen or Ti is the dominant process in degrading the oxide properties, not much is known about the time and length scales, both in plane and out of plane, at which this diffusion process occurs and the exact chemical composition at the metal-oxide interface. 

In this work, we image the influence of the metallic overlayers Ti and Au on the magnetic, structural and chemical properties of LMO thin films grown on STO substrates. The technique of scanning SQUID microscopy (SSM) is used, as it is a powerful tool to image local ferromagnetism with a micrometer-scale resolution \cite{Kirtley1999,Reith2017}. We observe a suppression of the ferromagnetism in LMO due to a Ti overlayer, a significant change in the out-of-plane lattice parameter up to several nanometers from the interface, corroborating the previous results on LSMO \cite{Homonnay2015}. Furthermore, we report on a time dependency of the suppression of ferromagnetism over a timescale of several days and we investigate in more detail the stoichiometry and valence states across the Ti-LMO interface. Finally, we show that by patterning the Ti layer we can locally modulate the magnetic properties and observe dipolar-like magnetic signals in sub-micrometer ferromagnetic structures.

\begin{figure}[h]
  \includegraphics[width = \linewidth]{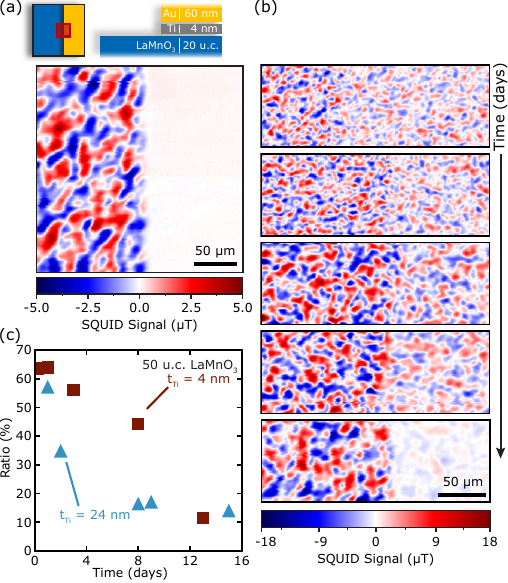}
  \caption{Scanning SQUID microscopy of LMO thin films partially covered by Ti and Au layers showing the suppression of the ferromagnetism on the Ti-covered side. (a) SSM scan of a 20 u.c. LMO thin film partially covered by Ti(4 nm)/Au(60 nm) layers, as illustrated by the stack on top of the graph. The shaded red area is an indication of the scan area during the measurement. (b) Panels of multiple SSM scans performed on the same 50 u.c. LMO thin film partially covered by Ti(4 nm)/Au(60 nm) over a time period of several days (0.2, 1, 3, 8 and 13 days from top to bottom respectively). (c) Ratio of the SQUID voltages on the covered- and uncovered sides as a function of time after deposition of the metallic layers for two 50 u.c. LMO samples with different thicknesses of Ti: 4 and 24 nm.}
  \label{fig1}
\end{figure}

For the experiments, metal layers of Ti and Au were sputtered in an inert Ar atmosphere on top of pulsed-laser deposited LMO on SrTiO$_3$ substrates. Details on the sample fabrication are provided in the supplemental materials \cite{supplemental}. In order to have a clear measurement of the influence of the metallic overlayers on LMO, Ti and Au were structured by means of photolithography and lift-off processing, and cover only one half of the LMO film. In this way it is possible to measure both covered and uncovered regions in the same SSM scan. The Au layer impedes the oxidation of the Ti film upon atmospheric exposure. Fig.~\ref{fig1}(a) shows a scan over the border region, where in the right side of the picture, a 20 unit cell (u.c.) thick LMO film (approximately 7.9 nm) is covered by 4 nm of Ti and 60 nm of Au. On the left (uncovered) side, the magnetic field exhibits a pattern in line with previous SSM experiments on ferromagnetic LMO thin films \cite{Wang2015, Reith2017}. It is presumed that the magnetism in the LMO is predominantly in-plane, with stray out-of-plane magnetic fields arising at domains walls. However, on the right side the signal is heavily suppressed, showing the effect of covering the LMO with the metal layers. Additional scans were performed entirely on the right side of the sample to show that no signal has actually been measured above the noise level of the SQUID. The SSM setup \cite{Kirtley1995,Kirtley1999,Wang2015} is operated at 4.2 K and has a sensitivity in the order of $\SI{10}{\micro\Phi_0\,Hz^{-1/2}}$ with a bandwidth of 1000 Hz, where $\Phi_0 = \SI{2e-15}{T\,m^2}$ is the magnetic flux quantum. For more details on the setup and SSM measurements see supplemental material Sec.~S3 \cite{supplemental}.

A control sample with no Ti layer, where only Au was sputtered using the exact same processing conditions as the Ti/Au samples, showed no significant decrease in the measured magnetic signal strength with respect to the uncovered LMO film, see Fig.~S3(a). Therefore, it is concluded that the presence of the overlaying Au layer does not significantly affect the ferromagnetism in LMO, nor should the sputtering process itself be responsible for the suppression observed. On the other hand, a sample with only a Ti layer did show a significant suppression, see Fig.~S3(b), so the Ti layer does play a dominant role in the effect, which is inline with previous work \cite{Homonnay2015}.

\begin{figure}[h]
  \includegraphics[width=\linewidth]{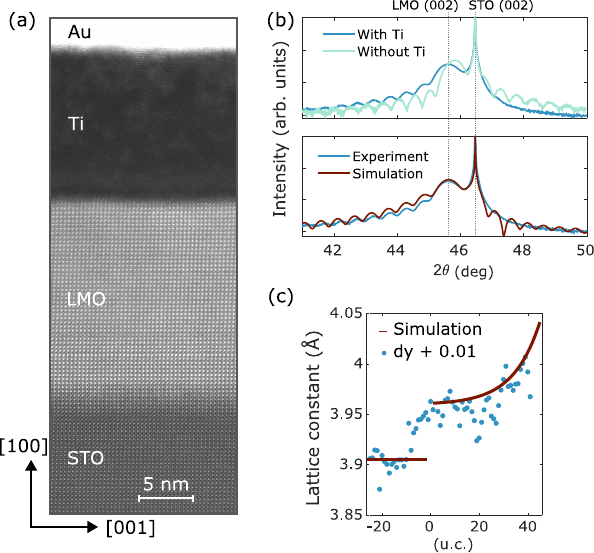}
  \caption{Structural characterization of the Ti/Au - LMO interface. (a) HAADF image of the STO/LMO/Ti/Au structure. (b) $2\theta-\omega$ scan of sample covered with and without Ti/Au and Experimental and simulated $2\theta-\omega$ scan of LMO covered with Ti/Au. (c) Extracted $c$ lattice parameter from the HAADF image as function of sample depth (blue circles). The $c$ axis lattice parameter as function of sample depth used in the model shown in (b) (red solid line).}
  \label{fig2}
\end{figure}

The ratio between the thickness of the LMO and Ti layers is important for the level of suppression of the ferromagnetic behavior. Experiments with a thicker LMO layer of 50 u.c. show that it is still possible to measure a significant stray magnetic field originating from the Ti-covered LMO for the same thickness of the Ti layer of 4 nm, although the suppression effect remains very clear. As before, a Au capping layer is used to impede oxidation of the Ti from the ambient environment. Fig.~\ref{fig1}(b) shows a panel comprised of several scans on a sample of 50 u.c.~thick LMO partially covered with 4 nm of Ti, performed at different moments in time after the deposition of the metal layers. Between the measurements the sample was stored at room temperature in a nitrogen flushed desiccator. As in Fig.~\ref{fig1}(a), the left side of the scans shows the uncovered region, while the right side shows the Ti/Au-covered film. To enable a quantitative comparison of the ferromagnetic signal in both regions of the scans, the variance (root mean square value) of the SQUID voltage signal (\textit{V}) in a defined area on each side of the border was calculated. The SQUID voltage is linearly dependent on the magnetic field as measured by the SQUID. This quantity can be used to estimate the variance of the strength of the measured magnetic field and can be used to compare between samples and when scanning parameters are changed \cite{Reith2017}. Then, for each scan, a ratio $R$ is defined by \textcolor{black}{$R = V_{\text{RMS,Ti}}/V_{\text{RMS,uncov.}}$}.
The panels in Fig.~\ref{fig1}(b) show that the suppression of the ferromagnetism in the Ti-covered LMO film increases with time on a timescale of several days. After 13 days, $R$ is reduced to approximately 12 \% compared to the uncovered region. It is also noteworthy that there must be at least two different rates of the decrease of $R$, because there is an abrupt decrease immediately after the deposition of the Ti layer, as after a few hours the ratio is already down to approximately 64 \%. Previous experimental studies dedicated to the oxidation of Ti thin films show that the oxidation process can be divided into two stages: a fast initial oxygen absorption step, taking place in a few minutes, followed by a slower process that can take hours to days \cite{Martin1994,Passeggi2002}. The occurrence of these different oxidation regimes is dependent on the Ti film thickness \cite{Vaquila1997}. Our experiments corroborate this. Thus, this leads us to believe that after depositing Ti/Au, oxygen vacancies are created rapidly, which is sufficient to completely suppress the magnetism in 20 u.c. thick films but not in the 50 u.c.~films. Here diffusion of the oxygen vacancies takes place on a longer time scale after the initial deposition, suppressing the ferromagnetic order further. Data on the diffusion of oxygen species in LMO is scarce, but recently a study gave indications for a high mobility of oxygen vacancies in LMO at moderate temperatures \cite{Rodriguez-Lamas2021}. \textcolor{black}{The observed change in SQUID signal strength on the uncovered \ce{LaMnO3} side and evolution of domain size in Fig.~\ref{fig2}(b) is a result of variations between different scanning SQUID measurements \cite{supplemental}}.

To study the dependence of the ferromagnetic suppression on the thickness of the Ti layer, several samples were fabricated while keeping the thickness of LMO to 50 u.c. and choosing different values of the Ti layer thickness. Fig.~\ref{fig1}(c) shows the ratio $R$ as a function of time for two different thickness values of the Ti layer: 4 and 24 nm. The comparison of the time evolution of the ratio for these two samples shows that for the same time period since the deposition of Ti, the level of suppression is different, \textit{i.e.} the Ti thickness seems to determine the rate of suppression of the ferromagnetism and possibly the final level of suppression. 

Theoretically, the STO substrate induces the correct amount of strain in the LMO film to stabilize the ferromagnetic ground state via strain-induced orbital ordering \cite{Hou2014}. In this model the two main prerequisites for ferromagnetism are: 1) crystallinity and 2) good stoichiometry, which is strongly tied to the oxygen content and Mn valency. This is reflected by studies showing the dependence of the magnetic order on the oxygen background pressure during growth \cite{Kim2010,Roqueta2015,Liu2019} and on the presence of a Ti overlayer causing migration of the O atoms in the LSMO film into the Ti layer \cite{Homonnay2015}. The latter study suggests a similar mechanism as in our case, however, the exact consequences on the valency and the composition at the metal-oxide interface is unknown. To get further insight into the mechanisms of the suppression, the two main prerequisites are investigated by means of scanning transmission electron microscopy (STEM), X-ray diffraction (XRD) and electron energy loss spectroscopy (EELS).

\begin{figure*}
    \centering
    \includegraphics[width = 1\textwidth]{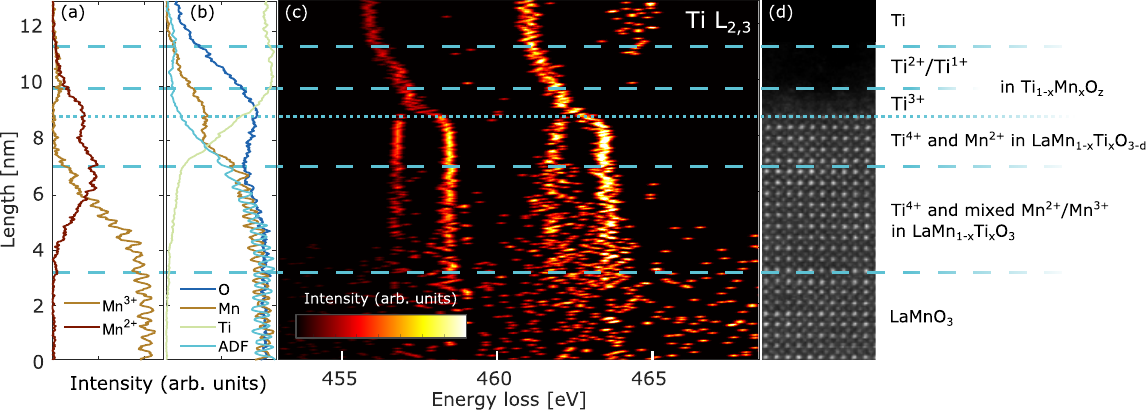}
    \caption{EELS data at the Ti L$_{2,3}$ edge with the corresponding ADF signal as function of sample thickness across the Ti/LaMnO$_{3}$ interface. The left 2-panels show the elemental composition as showed in the main text. The right panel the ADF image. From top to bottom the Ti valency evolves from elemental Ti, to Ti$^{1+}$/Ti$^{2+}$, to Ti$^{3+}$ and finally to Ti$^{4+}$ when the perovskite structure starts (it shares then the Mn B site of LaMnO$_{3}$). Also the tentative composition in each layer is indicated, where $x$ represents the minority substitutional element, $z$ the oxygen content of the amorphous transition metal oxide layer and $d$ the possible presence of oxygen vacancies.}
    \label{fig:Ti_valency}
\end{figure*}

Fig.~\ref{fig2}(a) shows a high-angle annular dark field (HAADF) STEM image of the STO/LMO/Ti/Au structure, where the LMO layer is approximately 50 u.c. thick, and the Ti and Au layers have a thickness of 12 and 60 nm, respectively. The image shows that the LMO film is grown epitaxially and crystalline on STO without any noticeable defects. 

The good crystallinity is confirmed by XRD. Fig.~\ref{fig2}(b) shows 2$\theta-\omega$ scans on a sample with and without a 24 nm-thick Ti layer (capped with Au). A clear STO 002 peak is visible in both scans accompanied by a LMO 002 peak with the corresponding Laue fringes. There is a significant difference in the LMO 002 peak position between the Ti-covered and uncovered samples, in accordance with \cite{Homonnay2015}. This indicates a (local) change of the out-of-plane lattice parameter $c$ in the LMO film, which is expected when oxygen vacancies are present \cite{Roqueta2015,Liu2019}. To quantify this change, the XRD pattern of a LMO film on a STO substrate was simulated, allowing for both the thickness of the LMO layer and the lattice parameter $c$ of the LMO to vary as function of film depth. This was performed with interactive XRD fitting \cite{Lichtensteiger2018}.  

The lower panel of Fig.~\ref{fig2}(b) shows the experimental XRD data of the Ti/Au covered area together with the XRD simulated pattern where $c$ varies as function of the sample depth with an exponential decay towards the STO/LMO interface. The red line in Fig.~\ref{fig2}(c) shows the variation of the lattice parameter that is used in the simulation. The best fit was obtained using an exponential dependence of the lattice parameter over a thickness of 45 LMO unit cells. Other models were also considered, where the lattice parameter $c$ is constant as function of sample depth or it has a step function-like behaviour, but these simulations deviated more from the experimental data. Fig.~\ref{fig2}(c) also shows the extracted difference between the lattice planes $dy$ as a function of sample depth from the HAADF image shown in Fig.~\ref{fig2}(a). Here a sample depth of 0 corresponds to the STO/LMO interface. The STO substrate and the LMO film can be distinguished by their different lattice parameters. A small upturn is observed towards the LMO/Ti interface, which suggest an increase of the out of plane lattice parameter towards the LMO/Ti interface.

The good crystal quality, significant change in the lattice parameter near the interface and previous studies all suggest that an off-stoichiometry due to oxygen vacancies is the cause of the suppression of the magnetism in our films. To get further insight in the valency states and exact composition, the LMO-Ti interface is studied with EELS. 

Fig.~\ref{fig:Ti_valency}(a) shows the EELS intensity of the Mn$^{2+}$ and Mn$^{3+}$ valency states and Fig. \ref{fig:Ti_valency}(b) shows the EELS intensity of the elements Mn, Ti and O as function of sample depth. The corresponding ADF image is shown in Fig. \ref{fig:Ti_valency}(d). The interface between the Ti and LMO is indicated with the bold blue line and the regions with different compositions across the interface are indicated by the dashed blue lines. The indicated composition reflects only the elements present in each layer where $x$ represents the minority substitutional element, $z$ the oxygen content of the amorphous transition metal oxide layer and $d$ the possible presence of oxygen vacancies. \textcolor{black}{The EELS signal of Ti is normalized to the signal in the Ti capping layer, while the EELS signal of the other elements is normalized to the signal in the LMO. Hence, the relative composition between the Ti and other elements cannot be deduced from this data}. From top to bottom the interface consist of elemental Ti, some form of Ti$_{x}$O$_{z}$, with a small amount of Mn doping, Ti \textcolor{black}{doped LaMn$_{1-x}$Ti$_x$O$_{3-d}$} and finally LaMnO$_{3}$. 

In the region of 3 nm above the interface Mn is present with a valency of Mn$^{2+}$. At the interface the presence of Ti drops rapidly, accompanied by a sudden increase of Mn. Interestingly, the oxygen content also drops for over a distance of 2 nm, after which it saturates. In the same region the Mn is predominately Mn$^{2+}$, which reflects the presence of oxygen vacancies. After the steep drop of the Ti signal, it decays slowly over a length of 5 nm to eventually vanish. Simultaneously, the valency of the Mn becomes completely Mn$^{3+}$. This shows that the interface region extends over approximately 8 nm and consist of Mn, Ti and O, with different valence states present. Focusing on the LMO, it is clear that for approximately 5 nm into the film from the interface the LMO is not stoichiometric due to a combination of oxygen scavenging and Ti interdiffusion. \textcolor{black}{Ti interdiffusion into the LMO has been observed before in studies at the STO-LMO interface \cite{Kaspar2018, Santolino20214, Guo2022}. However, in these works the length scale is several unit cells whereas we observe Ti diffusion over several nanometers. This difference could be caused firstly by an abundance of Ti at the sputtered LMO-Ti interface compared to the STO-LMO interface. Secondly a higher mobility caused by the non-stoichiometric interface and presence of oxygen vacancies in our samples can also lead to larger interdiffusion lengths}. 

Extra information about the evolution of the valence state of Ti across the interface can be obtained by looking at the Ti L$_{2,3}$ edge. Fig.~\ref{fig:Ti_valency}(c) shows the EELS data around the Ti L$_{2,3}$ edge with on the x-axis the energy, on the y-axis the sample depth and the color indicates the intensity. From the peak positions at the Ti L$_{2,3}$ edge the valency can be deduced \cite{Cao2002, Huang2010}. Far above the interface the L$_{2,3}$ peaks have a position of 461 eV and 455.8 eV, which would correspond to a Ti$^0$ valency \cite{Cao2002}. Just below the Ti-LMO interface four peaks are observed at positions of 461.9, 463.4 (L$_2$) and 456.7, 458.5 eV (L$_3$), which corresponds to a Ti$^{4+}$ valency \cite{Cao2002,Jang2011,Li2016}. In the transition from the Ti overlayer to the LMO film, the valence state of the Ti evolves smoothly from Ti$^0$ to Ti$^{4+}$ with intermediate regions of Ti$^{2+}$ and Ti$^{3+}$ \cite{Li2016}. 

Here, the STEM image shows that the LMO film is still fully crystalline, but the addition of a Ti scavenging layer leads to the active removal of the oxygen from the LMO layer, several nanometers in depth. This is shown by the presence of Ti$^{4+}$ (indicating the presence of Titanium on the Mn site in the LaMnO$_{3}$ perovskite structure) and Mn$^{2+}$ at the interface.

\begin{figure}[h]
  \includegraphics[width=\linewidth]{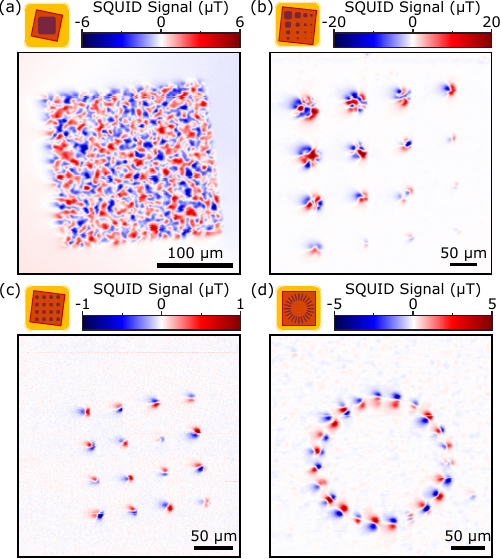}
    \caption{Scanning SQUID microscopy images of the patterned microstructures on 20 u.c. LMO by selectively covering with Ti and Au layers. In the top left of each subfigure an illustration is shown about the structures measured. (a) A $\SI{200}{} \times \SI{200}{\micro m^2}$ uncovered square. (b) Uncovered squares with descending size from top left to bottom right with a long side of respectively $\SI{25}{}$, $\SI{20}{}$, $\SI{15}{}$, $\SI{10}{}$, $\SI{5}{}$, $\SI{2.5}{}$, and $\SI{1}{\micro m}$. (c) Array of 16 uncovered $\SI{5}{} \times \SI{5}{\micro m^2}$ squares. (d) 23 bar structures disposed on a circumference, each with size $\SI{10}{} \times \SI{0.5}{\micro m^2}$.}
  \label{fig3}
\end{figure}

The suppression of the ferromagnetism in LMO achieved by the coverage with a Ti layer can be exploited to selectively pattern ferromagnetic structures in LMO thin films. To showcase this possibility, micrometer-sized areas of the LMO thin films were left uncovered by the Ti/Au bilayer, by means of lithography and lift-off. Four examples of such patterns can be seen in Fig.~\ref{fig3}. Areas where the SSM measures a distinct magnetic signal correspond to uncovered patches of the 20 u.c. LMO film, while in the regions covered by the metal layers the ferromagnetic signal is significantly suppressed. 

Fig.~\ref{fig3}(a) shows the SQUID signal of a $200\times\SI{200}{\micro m^2}$ square. The magnitude of the measured SQUID signal is similar to that of an uncovered film, see for example Fig.~\ref{fig1}(a). So, for patterns of this size, the size restrictions have no notable influence on the magnetism of the LMO underneath. Fig.~\ref{fig3}(b) shows that as the size of the structure decreases, the measured SQUID signal is reduced. The smaller SQUID signal can be explained by the fact that the SQUID resolution limits the signal and that by decreasing the structure size there are less ferromagnetic domains under the sensor area that contribute to the measured flux. The difference in the magnitude between the SQUID signal in Figs.~\ref{fig3}(a) and (b) is due to a different SQUID chip used in the measurements and a varying angle, for a more comprehensive explanation see the supplemental material \cite{supplemental}.

Interestingly, Fig.~\ref{fig3}(b) also reveals that as the structures get smaller, from squares with a side of $\SI{25}{\micro m}$ to $\SI{1}{\micro m}$, the magnetic behaviour changes and becomes reminiscent of magnetic multipoles and eventually dipoles. This can also be seen in Fig.~\ref{fig3}(c), where only squares with sides of $\SI{5}{\micro m}$ are imaged. This is the expected behaviour of a single ferromagnetic domain with in-plane magnetization \cite{Reith2017}. To lift the geometric symmetry of the squares, bars of $10\times\SI{0.5}{\micro m}$ were created and positioned radially inward on a circumference, as shown in Fig.~\ref{fig3}(d). Here, the dipole-like field distribution almost perfectly aligns radially, evidencing a favoured orientation of the magnetic moments due to the lateral constriction of the bars.

These results raise the question whether the small patches of LMO are uniformly polarized and represent real dipoles, or that the magnetic structure in the LMO is more complex and the magnetic field at the height of the pickup loop only emerges as that of a dipole. Domain sizes in LMO have been reported to be in the order of 200 nm \cite{Anahory2016, Gao2011}, which is an order smaller than our patterns. Our pickup loop scans at a height of approximately $\SI{3}{\micro m}$. Since this is an order of magnitude larger than the expected size of the domains, we don't expect to distinguish domains of these sizes individually. Finally, a uniform magnetically polarized patch of $5\times\SI{5}{\micro m^2}$ LMO would result in a magnetic field, at the pickup loop height, several orders of magnitude higher than we measure. So, although our small ferromagnetic patterns emerge as dipoles, we don't think that they are uniformly polarized. Rather we expect these LMO patches to posses more complex magnetic structures at length scales we can not resolve with our current experimental setup. 

In summary, scanning SQUID microscopy studies of LMO thin films partially covered by a thin Ti layer show the suppression of the ferromagnetism in LMO. For a combination of 20 u.c. of LMO and 4 nm of Ti, there is no stray magnetic field originating from the LMO film underneath. For thicker LMO films, the same thickness of Ti suppresses only partially the ferromagnetic phase in LMO and the rate of suppression is on the order of several days. STEM, EELS and XRD reveal the oxygen vacancies in the LMO layer near the interface, the complex elemental composition and the presence of different valence states across the interface. This is due to oxygen scavenging from the Ti layer and Ti interdiffusion for approximately 5 nm deep into the film.
This process first happens rapidly, resulting in immediate suppression in the 20 u.c. thick LMO samples. Further diffusion of oxygen vacancies deeper in the LMO films happens on timescales of days, resulting in a further suppression of ferromagnetism after initial Ti/Au deposition in 50 u.c. thick LMO samples. This suppression effect can be used to pattern the ferromagnetism in LMO thin films, as showcased by ferromagnetic (sub-)microstructures that were successfully fabricated in this way. Small structures show dipolar-like magnetic signatures, which raises questions about the effective domain size of the LMO and provides motivation for further research to study these structures with SQUID microscopes having higher spatial resolution.

The authors acknowledge support from the project "TOPCORE" (project number OCENW.GROOT.2019.048) which is financed by the Dutch Research Council (NWO). The authors acknowledge the research program “Materials for the Quantum Age” (QuMat) for financial support. This program (registration number 024.005.006) is part of the Gravitation program financed by the Dutch Ministry of Education, Culture and Science (OCW). JV acknowledges The eBEAM project which is supported by the European Union’s Horizon 2020 research and innovation programme FETPROACT-EIC-07-2020: emerging paradigms and communities. The Hercules fund 'Direct electron detector for soft matter TEM' from the Flemish Government.

\bibliography{References.bib}

\end{document}